\documentclass[journal]{IEEEtran}
\usepackage{amsmath,epsfig}
\usepackage{amssymb,float}
\usepackage{cite}
\usepackage{color}
\usepackage{bm}
\usepackage{fancyhdr}
\usepackage{multirow}
\newcommand{\thickhline}{%
    \noalign {\ifnum 0=`}\fi \hrule height 1pt
    \futurelet \reserved@a \@xhline}
\IEEEoverridecommandlockouts
\begin{document}
\title{Generalized SAT-Attack-Resistant \\Logic Locking}
\author{\IEEEauthorblockN{Jingbo Zhou and Xinmiao Zhang,~\IEEEmembership{Senior Member,~IEEE}}\\

\thanks{
This work is supported in part by US Air Force Research Laboratory under Award Number: FA8650-20-C-1719

The authors are with the The Ohio State University, Columbus, OH 43210, USA. Emails: \{zhou.2955, zhang.8952\}@osu.edu.}}

\IEEEtitleabstractindextext{%
\begin{abstract}
Logic locking is used to protect integrated circuits (ICs) from piracy and counterfeiting. An encrypted IC implements the correct function only when the right key is input. Many existing logic-locking methods are subject to the powerful satisfiability (SAT)-based attack. Recently, an Anti-SAT scheme has been developed. By adopting two complementary logic blocks that consist of AND/NAND trees, it makes the number of iterations needed by the SAT attack exponential to the number of input bits. Nevertheless, the Anti-SAT scheme is vulnerable to the later AppSAT and removal attacks. This paper proposes a generalized (G-)Anti-SAT scheme. Different from the Anti-SAT scheme, a variety of complementary or non-complementary functions can be adopted for the two blocks in our G-Anti-SAT scheme. The Anti-SAT scheme is just a special case of our proposed design. Our design can achieve higher output corruptibility, which is also tunable, so that better resistance to the AppSAT and removal attacks is achieved. Meanwhile, unlike existing AppSAT-resilient designs, our design does not sacrifice the resistance to the SAT attack.
\end{abstract}

}

\maketitle

\IEEEdisplaynontitleabstractindextext
\IEEEpeerreviewmaketitle

{\section{Introduction}\label{sec:introduction}}
\IEEEPARstart{N}{owadays}, integrated circuits (ICs) are designed and produced in a multi-vendor environment, which makes the designs face various security threats. In particular, netlists of the ICs may be obtained from reverse engineering or untrusted foundries. IP piracy and counterfeiting cause severe economic loss to the IC designers \cite{piracy, counterfeiting}. IC camouflaging \cite{Cocchi, Camouflaging1} resists reverse engineering \cite{reverse-engineering} by making functionally different logic gates look alike in the layout. However, it does not help in the case that the netlist is known. This paper focuses on developing a more secure logic-locking scheme. The basic idea of logic locking is to insert key-controlled logic components into the chip so that the chip does not function correctly without the right key.

Many logic-locking schemes have been developed previously by inserting XOR/XNOR gates \cite{Fault analysis,Yasin-logic,RajendranSecurity}, MUX gates \cite{PUF, Lee-logic}, or look-up tables (LUTs) \cite{Entanglement,LUT2} controlled by keys. However, these designs can be easily decrypted by the satisfiability (SAT)-based attack \cite{SAT}, which uses Boolean SAT solvers to iteratively update and solve the conjunctive normal form (CNF) formula of the target circuit. In each iteration, a distinguishing input pattern (DIP) is found, and the corresponding correct output is derived by querying the functioning chip. Then the correct output is utilized to exclude wrong keys. For many logic-locking schemes, only a small number of DIPs are needed to exclude all wrong keys. As a result, the SAT attack can be done in short time even if the key size is very large.

Several schemes have been proposed to resist the SAT attack \cite{XieMitigating,YasinSarlock}. The main idea is to adopt functional blocks that make the number of iterations and hence the query count in the SAT attack exponential. The Anti-SAT design \cite{XieMitigating} consists of two complementary function blocks implementing NAND/AND trees. The SAT attack excludes a disjoint set of wrong keys in each iteration and needs to go through all possible input patterns as DIPs before the right key is derived. In SARLock \cite{YasinSarlock}, the function blocks are designed so that each DIP can only exclude at most one wrong key. When the number of key bits is larger than the number of input bits, the SAT attack has to enumerate all input patterns. The Anti-SAT and SARLock schemes are vulnerable to more recent attacks. Due to the low corruptibility of all the wrong keys in those schemes, they are subject to the AppSAT attack \cite{AppSAT}. Additionally, the AND/NAND functions in the two blocks of the Anti-SAT scheme lead to large signal skew, which makes this scheme also subject to the signal probability skew (SPS) attack \cite{Removal}.

Combining a traditional high output error scheme, such as \cite{Fault analysis,Yasin-logic,PUF}, with a SAT-resilience block, such as the Anti-SAT or SARLock, can increase the overall output corruptibility. However, the compound scheme can be reduced to standalone SAT-resilient block, from which the AppSAT attack \cite{AppSAT} can recover an approximate key. Also, the AppSAT-guided removal (AGR) attack \cite{Removal} and bit-flip attack \cite{bit-flip} can separate the key inputs for high error output and the key inputs for the SAT-resistant block. Then using structural analysis, the output signal of the SAT-resilient block can be identified and set to constant to make the circuit function correctly.

It was claimed in \cite{AppSAT,humble} that there is a fundamental trade-off between the corruptibility of a logic-locking block and number of queries needed in the SAT attack. Increasing the corruptibility will always reduce the query count. The stripped-functionality logic locking (SFLL) \cite{SFLL} increases the corruptibility by adopting Hamming distance checkers of higher weight. The diversified tree logic (DTL) \cite{AppSAT} and its special case, error-controllable encryption (ECE) \cite{ECE}, replace the AND gates in an AND tree structure by XOR/OR/NAND gates to control the true set of the tree output signal. The DTL can be also incorporated in the Anti-SAT \cite{XieMitigating} and SARLock \cite{YasinSarlock} schemes. All of these schemes increase the corruptibility of each wrong key at the cost of reduced query count. Besides, the SFLL is subject to the functionality analysis attack \cite{FALL} and the structural analysis attack \cite{Fangfei}, both of which try to analyze the crucial components in the SFLL to get the right keys.

This paper proposes a generalized (G-) Anti-SAT scheme. The proposed design not only resists the SAT attack, but also improves the approximate resiliency without sacrificing the query count. Besides, it has better resistance to removal attacks. Different from the previous schemes, the wrong key set that can be excluded by each DIP in our design has a unique wrong key that is not included in the wrong key sets of any other DIPs. This property makes the design have exponential query count and hence always resistant to the SAT attack. At the same time, the wrong key sets can have overlaps in order to increase the corruptibility. Compared to the Anti-SAT scheme \cite{XieMitigating}, our proposed design is generalized in two dimensions. The two functions do not have to be AND/NAND or complementary. The Anti-SAT design is just a special case of our proposed G-Anti-SAT scheme. {The cascaded(CAS)-lock logic-locking block \cite{CAS} that is implemented with cascaded AND/OR gates can also increase the corruptibility without sacrificing the resistance to the SAT attack. It is also a special case of our proposed design.}

The major contributions of this paper are as follows.

\begin{enumerate}
\item Generalized constraints on the two function blocks are proposed to make the number of SAT attack queries exponential to the input size.
\item Following the constraints, logic-locking blocks can be designed to increase the corruptibility without sacrificing the query count. Design procedures using K-maps are given for the G-Anti-SAT block to achieve higher corruptibility and at the same time exponential query count.
\item The generalized constraints allow a variety of functions to be used in the logic-locking block. The functions do not have to be AND/NAND or complementary. The variations of functions allow better resistance to attacks based on structural or function analysis.
\item The non-AND/NAND function and higher corruptibility of our design allow better resistance to the removal attacks \cite{Removal}, bypass attack \cite{Bypass} and bit-flip attack \cite{bit-flip}.
\end{enumerate}

This paper is organized as follows. Section II briefly introduces available attacks and the Anti-SAT design. Section III proposes our G-Anti-SAT constraints. Section IV presents methodologies for developing functions satisfying the G-Anti-SAT constraints using K-maps. Analysis and experimental results showing the resistance of our design to various attacks are given in Section V. Discussions and conclusions follow in Section VI and Section VII, respectively.

\section{Background}
This section introduces basic knowledge about the SAT attack, Anti-SAT block, AppSAT and removal attacks.

\subsection{SAT attack}
The SAT attack \cite{SAT} is a powerful technique against logic-locking. The attack model assumes that the attacker has access to the gate-level netlist of the locked circuit, which can be obtained by reverse engineering or from an un-trusted foundry. Represent the netlist of the locked circuit by $Y = f_e(X, K)$, where $X$, $K$, and $Y$ are the primary input, key input, and primary output vectors, respectively. Its CNF formula is represented as $C_e(X, K, Y)$. It is also assumed that the attacker has an activated chip with the right key input, whose function is denoted by  $Y = f_o(X)$. Queries can be made on the chip to find the correct outputs for given inputs.

The SAT attack finds the right key by excluding all wrong keys through utilizing DIPs. Initially, a SAT solver is applied to the following formula
\begin{equation}\label{first iter}
F_0 := C_e(X, K_1, Y_1) \land C_e(X, K_2, Y_2) \land (Y_1 \neq Y_2)
\end{equation}
to solve for an $X$ that leads to different outputs, $Y_1$ and $Y_2$, under two different keys, $K_1$ and $K_2$. This $X$ is referred to as a DIP and is denoted by $X_1^d$. Then the activated chip is queried to get the corresponding correct output $Y_1^d = f_o(X_1^d)$. After that, new constraints according to $X_1^d$ and $Y_1^d$ are added and the SAT formula in \eqref{first iter} is updated as $ F_1 = F_0 \land C_e(X_1^d, K_1, Y_1^d) \land C_e(X_1^d, K_2, Y_1^d)$. Then the updated SAT formula is solved for another DIP $X_2^d$, which is used to query the activated chip and accordingly update the SAT formula. This process is repeated iteratively. In the \textit{i}th iteration, the SAT formula is
\begin{equation*}
  F_i = F_0 \bigwedge_{j=1}^{i}(C_e(X_j^d, K_1, Y_j^d)\land C_e(X_j^d, K_2, Y_j^d)).
\end{equation*}
If $F_i$ is satisfiable, then there exist at least one pair of keys $K_1, K_2$, and $X_{i+1}^d$ such that $f_e(X_{i+1}^d, K_1) \neq f_e(X_{i+1}^d, K_2)$, which means not all wrong keys have been excluded from the key space. When the SAT formula is no longer satisfiable in an iteration, say $\lambda$, the algorithm stops. At this time, the right key can be derived by solving the following SAT formula
\begin{equation*}
  F := \bigwedge_{i=1}^{\lambda}C_e(X_i^d, K, Y_i^d).
\end{equation*}

\subsection{Anti-SAT block}
\begin{figure}[t]
  \centering
  \includegraphics[width=3.5in]{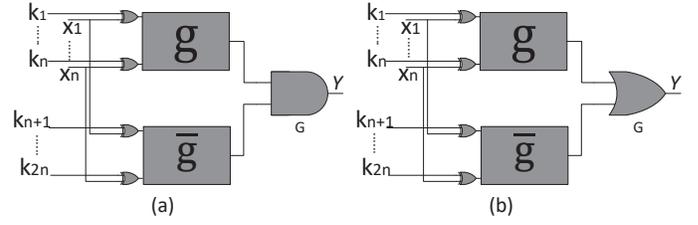}
  \caption{Anti-SAT block. (a) Type-0 block; (b) Type-1 block}
\label{Anti-SAT drawing}
\end{figure}

The Anti-SAT block is proposed in \cite{XieMitigating}. It is composed of two complementary functions $g$ and $\bar g$ as shown in Fig. \ref{Anti-SAT drawing}. These two functions share the same input $X$ but have different keys. The outputs of the two functions can be ANDed or ORed to generate the output as shown in Fig. \ref{Anti-SAT drawing} (a) and (b), respectively. They are referred to as the type-0 and type-1 blocks, respectively. Let $K_g = [k_1,k_2,\cdots, k_{n}]$ and $K_{\bar g} = [k_{n+1},k_{n+2},\cdots,k_{2n}]$. Any $K_g = K_{\bar g}$ are right keys for the Anti-SAT scheme. Since $g$ and $\bar g$ are complementary functions, the correct output of the type-0 block in Fig. \ref{Anti-SAT drawing}(a) is '0', and that of the type-1 block in Fig. \ref{Anti-SAT drawing}(b) is '1'.

Let $X=[x_1,x_2,\cdots, x_{n}]$. The input to the $g$ function in Fig. \ref{Anti-SAT drawing}(a) is $L=X\oplus K_g$. Define
\begin{equation}\label{LTLFp}
\begin{aligned}
  &\textbf{G}^T = \{L|g(L) = 1\},\hspace{0.5cm} (|\textbf{G}^T|=p)\\
  &\textbf{G}^F = \{L|g(L) = 0\},\hspace{0.5cm} (|\textbf{G}^F| = 2^n - p)
\end{aligned}
\end{equation}
In the remainder of this paper, $\textbf{G}^T$ is referred to as the true set. In \cite{XieMitigating}, it has been derived that the total number of iterations needed by the SAT attack on the structure shown in Fig. \ref{Anti-SAT drawing} is lower-bounded by
\begin{equation*}\label{Xie proof}
  \lambda \geq \frac{2^{2n} - 2^n}{p \times (2^n - p)} .
\end{equation*}
When $p=1$ or $2^n-1$, $\lambda\geq 2^n$. Since there are $2^n$ input patterns, this means that the number of iterations needed by the SAT attack and hence the number of queries to the chip is $2^n$ and all input patterns need to be gone through as DIPs to find the right keys. In this case, the SAT attack is effectively resisted. A natural candidate for $g$ that satisfies $p=1$ or $p=2^n-1$ is the AND or NAND of all inputs.

\subsection{AppSAT attack}
Logic-locking schemes with low corruptibility can be decrypted by the AppSAT attack \cite{AppSAT}, which avoids exponential number of iterations by introducing random query reinforcement and stopping the query process early. {Corruptibility is defined as the number of input patterns making the output wrong under wrong keys \cite{AppSAT}}  After every certain number of iterations in the SAT attack, random input patterns are utilized to query the activated chip, and the constraints from the queries are added to the CNF formula. If the output has low corruptibility, the portion of the input queries that generate the wrong output falls below a threshold. If this happens for a number of rounds, the algorithm terminates and returns an approximate key.

When the $g$ and $\overline{g}$ in the Anti-SAT block of Fig. \ref{Anti-SAT drawing} are $n$-input AND and NAND gates, respectively, there is only one input pattern that makes the output wrong for each wrong key. Such an Anti-SAT design has low corruptibility and is subject to the AppSAT attack. To address this issue, it was proposed in \cite{XieMitigating} to combine the Anti-SAT block with traditional logic-locking schemes to increase the overall output corruptibility. However, the AppSAT attack can reduce this compound scheme to standalone SAT-resilient scheme, from which the keys can be recovered. The corruptibility has been increased in the SFLL \cite{SFLL}, DTL \cite{AppSAT}, and ECE \cite{ECE} schemes. Nevertheless, these designs lead to reduced number of iterations and hence query count in the SAT attack. There was a trade-off between the corruptibility and the query count in the previous designs.

\subsection{Removal attack}
Removal attacks \cite{Removal} can be utilized to identify the last gate of the logic-locking block, such as the gate $G$ of the Anti-SAT block in Fig. \ref{Anti-SAT drawing}. Then the output of this gate can be replaced by the correct signal to make the circuit function correct. For the type-0 and type-1 Anti-SAT blocks, the correct outputs are `0' and `1', respectively.

The removal attack for the Anti-SAT block can be carried out using the signal probability skew SPS, which is defined as
\begin{equation*}
  s_x = Pr[x = 1] - 0.5
\end{equation*}
for a signal $x$. Since $0 \leq Pr[x=1] \leq 1$, the range of $s_x$ is $[-0.5, 0.5]$. For a logic gate with two inputs whose SPS values are $s_1$ and $s_2$, its absolute difference (ADS) value is defined as
\begin{equation*}
 ADS= |s_{1} - s_{2}|.
\end{equation*}

Assuming that $X,K_1, K_2$ are random. Then the SPS values of the inputs to the XOR gates in the Anti-SAT block are zero. Accordingly, the outputs of the XOR gates have zero SPS values. The SPS value for the output of an $n$-input AND gate is calculated as $s_{n-AND} = \prod_{i=1}^n(0.5+s_i)-0.5$, where $s_i$ is the SPS of the $i^{th}$ input. Since $s_i=0$ for the AND gate in the $g$ function, $s_{g(X,K_1)} = 0.5^n - 0.5$. As $n \to \infty$, $s_{g(X,K_1)} \approx -0.5$. Similarly, for the $n$-input NAND gate output from $\overline{g}$, the SPS is $s_{\overline{g}(X,K_2)} = 0.5 - 0.5^n$, which approaches $0.5$ for large $n$. As a result, for the last gate, $G$, of the Anti-SAT block in Fig. \ref{Anti-SAT drawing}(a), the output SPS is -0.5 and its ADS is $ADS_G = |s_{g(X,K_1)} - s_{\overline{g}(X,K_2)}|\approx 1$, if the number of inputs to the Anti-SAT block is large.

It was found in \cite{Removal} that the ADS values for the gates in a circuit are rarely very high. Hence the $G$ gate may be identified by first sorting out the gates with the highest $ADS$ values. In the case that there are multiple candidates whose $ADS$ values are very close, the transitive fan-in (TFI) of the candidate gates are analyzed. The TFI traces back the inputs of the candidate gates and finds how many key bits contribute to the inputs. The $G$ gate should have all $2n$ key bits as contributors. Once the $G$ gate is identified, its output signal can be replaced by `0' or `1' in the circuit when its SPS is negative or positive, respectively. This attack is named as the SPS attack.

In order to resist the SPS attack, structural obfuscation can be applied to the Anti-SAT scheme. After applying additional keys to obfuscate the structure, the SPS attack cannot detect the final gate $G$ by simply sorting the ADS values. However, it was also mentioned in \cite{Removal} that the removal and AppSAT attacks can be combined, which is termed as the AppSAT guided removal (AGR) attack. The combined attack uses the AppSAT attack to separate the keys for structural obfuscation from the keys to the Anti-SAT block. This is possible since the approximate values of the key inputs to the Anti-SAT block returned by the AppSAT attack over the iterations fluctuate due to the low corruptibility. After that, structural analyses can help the attacker find the final gate $G$. Then, similarly, the output signal of $G$ is replaced by the correct value decided according to the SPS value.

\section{Generalized Anti-SAT Constraints}
The main reason that the Anti-SAT block is subject to the AppSAT and removal attacks is that $p$, the cardinality of the true set of the function $g$, is either too small or too big and the two functions are complement of each other. As a result, the corruptibility of the output is very low. On the other hand, such $p$ is needed in the Anti-SAT design to maximize the number of iterations in the SAT attack. To solve this dilemma, true sets that have medium cardinality and at the same time lead to maximum SAT attack iterations are necessary. In this section, generalized constraints on the true sets for resisting the SAT attack are proposed. Our generalization allows a wide range of true set cardinality. Accordingly, logic-locking blocks with higher corruptibility can be designed to achieve better resilience to the AppSAT and removal attacks and resist the SAT attack at the same time. For convenience, some notations used in the following discussions are listed in Table \ref{Notation}. In the remainder of this paper, bold capital math symbols, such as $\mathbf X_n$ and $\mathbf F^F$, denote sets and capital symbols in normal font, such as $X$ and $F^F$ represent vectors.

\begin{table}
  \renewcommand\arraystretch{1.3}
  \centering
  \caption{Summary of notations}
  \label{Notation}
  \begin{tabular}{l|l}
  \hline
  $L$ & The input vector to function $f$ or $g$ \\
  $X$ & The data input vector to the logic-locking block\\
  $\lambda$ & The number of iterations needed in the SAT attack\\
  $K_f$ & The key input vector to block $f$ \\
  $\textbf{X}_n$ & The set of all $n$-bit input vectors \\
  $\textbf{F}^T$ & True set of function $f$ \\
  $\textbf{F}^F$ & False set of function $f$ \\
  $\textbf{WK}_{X}$ & The set of wrong keys that input $X$ can exclude \\
  $\textbf{F}^F_{K_f}$ & The set of all vectors in $\textbf{F}^F$ XORed with $K_f$ \\

  $\textbf{D}_{S-\textbf{S}}$ & $\{D| D=S \oplus S_i, \forall S_i \in \textbf{S}/S\}.$\\
  $\textbf{D}_{\textbf{S}}$ & $\{D|D = S_1 \oplus S_2, \forall S_1 \neq S_2 \in
  \textbf{S}\}$\\
    $\oplus$ & Bit-wise XOR operation \\
  $||$ & Concatenation operation \\
  $\&$ & Logic AND operation \\
  $+$ & Logic OR operation \\
  \hline
  \end{tabular}
\end{table}

\subsection{Wrong key sets analysis}
\begin{figure*}
    \centering
    \includegraphics[width=6in]{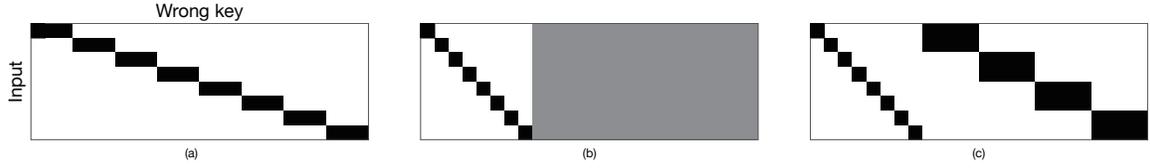}
    \caption{Input-wrong key array, where a black block indicates wrong output for the corresponding input pattern and wrong key. (a) array for the Anti-SAT design; (b) array for the relaxed wrong key set constraint; (c) example array satisfying the relaxed wrong key set constraint}
    \label{wrongkey example}
\end{figure*}
Define $\textbf{WK}_{X}$ as the wrong key set that input vector $X$ can exclude. In other words, $\textbf{WK}_{X} = \{{WK}|f_e({WK}, X) \neq f_o(X)\}$. If the wrong key set of an input vector has already been covered by the wrong key sets of the DIPs in the previous iterations of the SAT attack, then this vector will not be selected as a DIP in the rest of the SAT attack. Therefore, if a circuit can be decrypted by the SAT attack in a limited number of iterations, there must be many input vectors whose wrong key sets cover each other and are not selected as DIPs.

Consider a circuit that has $n$ inputs and requires $\lambda$ iterations in the SAT attack. Denote the set of DIPs by $\textbf{X}_{DIP}$. Accordingly,  $\lambda = |\textbf{X}_{DIP}|$. If a block needs to be resistant to the SAT attack, $\lambda$ needs to be as big as possible, which is $2^n$. This means that each possible $n$-bit input pattern can exclude some unique wrong keys that the other inputs cannot exclude. Take a 4-bit-input type-0 Anti-SAT block as an example. When $p$ in \eqref{LTLFp} is $1$, for each possible input $X$, $|\textbf{WK}_{X}| = 15$. The key input has 8 bits and hence $2^8$ different patterns. From \cite{XieMitigating}, 16 of the keys are correct. Hence, the total number of wrong keys is $2^8-16=240$. Therefore, the number of DIPs and the number of iterations carried out by the SAT attack should be $\lambda \geq \frac{240}{15} = 16$. On the other hand, for 4-bit input, there are $2^4$ patterns. Hence, $\lambda=16$. Apparently, for this Anti-SAT scheme, the wrong key sets for different input patterns do not have any overlap. In other words,
\begin{equation}\label{constraint Anti-SAT}
  \forall X_1 \neq X_2 \in \textbf{X}_n,\hspace{0.2cm}\textbf{WK}_{X_1} \cap \textbf{WK}_{X_2} = \varnothing,
\end{equation}
where $\textbf{X}_n$ is the set of all possible input patterns of $n$ bits. {The wrong key sets of the Anti-SAT design can be illustrated by the array in Fig. \ref{wrongkey example} (a), in which the rows and columns represent all input patterns and wrong keys, respectively, and the black blocks in the row for input pattern $X$ indicate $\textbf{WK}_X$. It is clear in this figure that the wrong key sets for different input patterns do not overlap.}

$\lambda$ can still be made equal to $2^n$ to be resistant to the SAT attack even if there are overlaps among the wrong key sets. The Anti-SAT block is a special case. In addition, the functions of the two blocks do not have to be complementary of each other as in the Anti-SAT block.

\subsection{Highlights of proposed G-Anti-SAT block}
Our proposed G-Anti-SAT scheme generalizes the previous approach by allowing the wrong key sets of different input patterns to have overlaps. Instead of \eqref{constraint Anti-SAT}, our design requires that
\begin{equation}\label{resist SAT}{\small
\forall X_1 \neq X_2 \in \textbf{X}_n, \exists K \,s.t.\, (K\in \textbf{WK}_{X_1}) \& (K\notin \textbf{WK}_{X_2})}.
\end{equation}
In other words, each wrong key set has at least one distinct element. {This constraint is illustrated in the input-wrong key array in Fig. \ref{wrongkey example} (b). The distinct elements are denoted by the black cells in the diagonal. However, the rest of the input-wrong key array, as represented by the gray area, can have any patterns. Unlike that of the Anti-SAT design in Fig. \ref{wrongkey example} (a), the wrong key sets corresponding to different input patterns can have overlaps}. Adopting the relaxation in \eqref{resist SAT}, there are still $\lambda=2^n$ DIPs. Hence, our generalized design is still resistant to the SAT attack. By allowing overlapping wrong key sets, the cardinality of the true set is relaxed so that it can be integers other than 1 or $2^n-1$. The distinct element in each wrong key set is a wrong key that has only one input leading to the wrong output. However, by choosing functions with different true set cardinalities, the other wrong keys can have more inputs leading to the wrong output and hence higher corruptibility as shown by the example in Fig. \ref{wrongkey example} (c). As a result, better resiliency towards the AppSAT and removal attacks is achievable.

A second dimension of generalization is done in our design by allowing the two functions to be $f$ and $g$, which are not necessarily complementary of each other. This enables larger design space and makes it even more difficult for the attacker to guess the functions adopted in the logic-locking block. Our non-complementary designs also have improved corruptibility and have better resilience towards the AppSAT and removal attacks.

In the following, subsection III.C.1) analyzes the constraints on the true sets to satisfy \eqref{resist SAT}. When non-complementary blocks are adopted, it is non-trivial to identify the right keys. The constraint to ensure the existence of right keys is provided in subsection III.C.2). Section IV presents construction methods for the true sets by using K-maps. K-maps help to not only highlight the constraints need to be satisfied but also facilitate the design of the functions to achieve given true set cardinality.

\subsection{Constraints for SAT attack resistance and right key existence}
\subsubsection{Constraints for resisting SAT attack}
\begin{figure}
  \centering
  \includegraphics[width=2.4in]{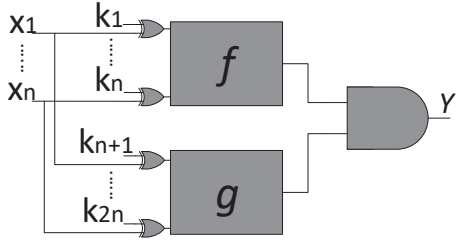}
  \caption{Architecture of the proposed type-0 G-Anti-SAT block}
  \label{alternative structure}
\end{figure}

Fig. \ref{alternative structure} shows our proposed G-Anti-SAT block for type-0 design. Different from the Anti-SAT design \cite{XieMitigating}, the functions of the two blocks do not have to be complementary of each other. Similarly, the last gate can be replaced by an OR gate to be a type-1 design. In the following, analysis is carried out on the type-0 design shown in Fig. \ref{alternative structure}. All the proposed analysis and constraints can be extended easily for the type-1 design.

Similarly, define
\begin{equation*}
\begin{aligned}
  \textbf{F}^T &= \{L|f(L) = 1\}\hspace{0.3cm} \textbf{F}^F = \{L|f(L) = 0\}.\\
\end{aligned}
\end{equation*}
Following the convention in \cite{XieMitigating}, `1' is considered as the incorrect output for a type-0 block. For the architecture in Fig. \ref{alternative structure}, the output function is $y = f(X \oplus K_f) {\&} g(X \oplus K_g)$, where $K_f$ and $K_g$ are the key inputs of block $f$ and $g$, respectively. $y$=`1' only if $F^T= X \oplus K_f  \in \textbf{F}^T$ and $ G^T = X \oplus K_g \in \textbf{G}^T$. Therefore, the wrong key patterns in $\textbf{WK}_{X}$ are in the format of $[X \oplus F^T || X \oplus G^T]$, where $||$ means concatenation. Accordingly, \eqref{resist SAT} can be interpreted as
\begin{equation}\label{proof1}
  \begin{aligned}
    &\forall X_1 \neq X_2 \in \textbf{X}_n, \exists F_1^T \neq F_2^T \in \textbf{F}^T, G_1^T \neq G_2^T \in \textbf{G}^T \\
    &s.t. \hspace{0.2cm}[F_1^T \oplus X_1 \hspace{0.1cm} || \hspace{0.1cm} G_1^T \oplus X_1] \neq [F_2^T \oplus X_2 \hspace{0.1cm} ||\hspace{0.1cm} G_2^T \oplus X_2].
  \end{aligned}
\end{equation}
Since $\textbf{X}_n$ includes all $n$-bit vectors, for any $F_1^T \neq F_2^T$, there must exists $X\in \textbf{X}_n$ such that $F_1^T \oplus F_2^T = X$. $X$ can be also rewritten as the sum of two elements in $\textbf{X}_n$, {\it i.e.} $X = X_1 \oplus X_2$. Hence the constraint $\forall X_1 \neq X_2$, $F_1^T \oplus X_1  \neq F_2^T \oplus X_2$ can never be satisfied. Similarly, there do not exist $G_1^T \neq G_2^T $ such that $G_1^T \oplus X_1  \neq G_2^T \oplus X_2$,\ $\forall X_1 \neq X_2$. Therefore, to satisfy the constraints in \eqref{proof1}, $\textbf{F}^T$ and $\textbf{G}^T$ need to be designed jointly so that $F_1^T \oplus X_1 =F_2^T \oplus X_2$ and $G_1^T \oplus X_1= G_2^T \oplus X_2$ are not true at the same time.

Define the binary distance between two vectors $X_1$ and $X_2$ as $X_1 \oplus X_2$. Let $\textbf{D}_{S-\textbf{S}}$ be a set consisting of binary distances between an element $S\in \textbf{S}$ and all the other elements in $\textbf{S}$. In other words,
\begin{equation}\label{distance structure}
  \textbf{D}_{S-\textbf{S}} = \{D| D=S \oplus S_i, \forall S_i \in \textbf{S}/S\}.
\end{equation}
Then $\textbf{D}_{F^T-\textbf{F}^T}$ is the set of vectors consisting of $F^T\oplus F_1^T$ for every $F_1^T\in \textbf{F}^T$ and $F_1^T\neq F^T$. If $F^T \oplus X_1 =F_1^T \oplus X_2$ is satisfied, $F^T \oplus F_1^T= X_1 \oplus X_2$. Hence, $X_1\oplus X_2$ is also in the set $\textbf{D}_{F^T-\textbf{F}^T}$. Similarly, the $X_1\oplus X_2$ of the $X_1$ and $X_2$ satisfying the constraint that $G_1^T \oplus X_1= G_2^T \oplus X_2$ is in the set $\textbf{D}_{G^T-\textbf{G}^T}$. Accordingly, the constraints in \eqref{proof1} are translated to
\begin{equation}\label{criteria1}
\begin{split}
{\text {\bf Constraint 1:}}\ \exists F^T&\in\textbf{F}^T, G^T\in \textbf{G}^T \\
&s.t. \, \textbf{D}_{F^T-\textbf{F}^T} \cap \textbf{D}_{G^T-\textbf{G}^T} = \varnothing.
\end{split}
\end{equation}
The above equation gives a constraint equivalent to that in \eqref{resist SAT}. However, this constraint can be utilized to construct $\textbf{F}^T$ and $\textbf{G}^T$ more easily.

From Constraint 1, it is clear that the functions $f$ and $g$ do not have to be complementary, and $|\textbf{F}^T|$, $|\textbf{G}^T|$ do not need to be $1$ or $2^n-1$. Therefore, the $f$ and $g$ blocks do not need to be AND and NAND gates, respectively, as in the Anti-SAT block \cite{XieMitigating}. The Anti-SAT block is just a special case of our proposed design. Many different functions can be chosen for $f$ and $g$. In the design of $f$ and $g$, an arbitrary set can be chosen as $\textbf{F}^T$ first. For the selected $\textbf{F}^T$, the choice of $\textbf{G}^T$ may not be unique. Any $\textbf{G}^T$ satisfying Constraint 1 can be utilized to achieve SAT-attack resilience.

{\bf Example 1} Take the structure in Fig. \ref{alternative structure} with 4-bit input as an example. Different from the previous design, the $f$ and $g$ functions are allowed to be non-complementary. First, let $\textbf{F}^T = \{0,1,2,3\}$. To simplify the notations, decimal numbers are used to represent vectors here. For example, 2 represents the binary vector $[0, 0, 1, 0]$. It turns out $\{0,8,9,11,10\}$ is one of the possible sets for $\textbf{G}^T$ that can satisfy Constraint 1. When $F^T = G^T = 0$, the two sets in Constraint 1 are disjoint. For a given $\textbf{F}^T$, the corresponding $\textbf{G}^T$, $F^T$, and $G^T$ satisfying Constraint 1 can be found easily using K-maps and the procedure will be detailed in Section IV. $\textbf{F}^T$ and $\textbf{G}^T$ are the minterm numbers of the $f$ and $g$ functions, respectively. The logic formula for $f$ and $g$ can be derived accordingly. For the above choice of $\textbf{F}^T$ and $\textbf{G}^T$, $f(L) = \overline{l_3} \& \overline{l_2}$ and $g(L) = l_3 \& \overline{l_2} + \overline{l_2} \& \overline{l_1} \& \overline{l_0}$,
where $L=[l_3,l_2, l_1, l_0]$ is the 4-bit input to $f$ and $g$ and `$\&$' and `$+$' denote the logic AND and OR operations, respectively.

{\bf Example 2} The $f$ and $g$ in our design can be complementary as well. Select $\textbf{F}^T = \{6,8,9,10,11,12,13,14,15\}$ and $\textbf{G}^T = \{0,1,2,3,4,5,7\}$. It can be found that $F^T = 5$ and $G^T = 6$ satisfy Constraint 1. The corresponding functions $f$ and $g$ are $f(L) = l_3 + l_2 \& l_1 \& \overline{l_0}$ and $g(L) = \overline{f(L)}$.

{\bf Example 3} Constraint 1 is not sufficient to guarantee the existence of right keys. For example, take $\textbf{F}^T = \{0,1,2,3\}$ and $\textbf{G}^T = \{0,4,8,12\}$. Constraint 1 is also satisfied by taking $F^T = G^T = 0$. Accordingly, $f(L) = \overline{l_3} \& \overline{l_2}$ and $g(L) = \overline{l_1} \& \overline{l_0}$. However, in this case, from exhaustive search, there does not exist a right key $K^*$ such that $f_e(X, K^*) = f_o(X)$ for every possible $X$.


\subsubsection{Constraints for existence of right keys}

When $f$ and $g$ are not complementary, additional constraints need to be introduced to guarantee the existence of right keys.

The correct output of the type-0 logic-locking block in Fig. \ref{alternative structure} is `0'. Hence, right keys are $[K_f \hspace{0.05cm}||\hspace{0.05cm}K_g]$ such that for every $X\in \textbf{X}_n$, $f(X\oplus K_f) = 0$ or $g(X\oplus K_g) = 0$. Define $\textbf{F}^F_{K_f} = \{X|X = F^F \oplus K_f,\ \forall F^F \in \textbf{F}^F \}$. It is the set of $X$ that makes $f(X\oplus K_f) = 0$. Similarly, $\textbf{G}^F_{K_g} = \{X|X = G^F \oplus K_g,\  \forall G^F \in \textbf{G}^F\}$ is the set of $X$ that makes $g(X\oplus K_g) = 0$. Therefore, the right keys $[K_f \hspace{0.05cm}||\hspace{0.05cm}K_g]$ should satisfy
\begin{equation}\label{criteria3}
  (\textbf{F}^F_{K_f}\cup \textbf{G}^F_{K_g} ) = \textbf{X}_n.
\end{equation}

For a selected function $f$, a function $g$ can be designed to satisfy \eqref{criteria3}. From the definition, $\textbf{F}^T\cup \textbf{F}^F=\textbf{X}_n$ and  $\textbf{F}^T\cap \textbf{F}^F=\varnothing$. Hence, for any $K_f$, $\textbf{F}^F_{K_f} \cup \textbf{F}^T_{K_f} = \textbf{X}_n$. In the case that $f$ and $g$ are not complementary, if $\textbf{G}^F_{K_g}\supseteq  \textbf{F}^T_{K_f}$, then \eqref{criteria3} would be satisfied. Define the binary distance structure of a set $\textbf{S}$ as
\begin{equation}\label{Ds}
  \textbf{D}_{\textbf{S}} = \{D|D = S_1 \oplus S_2, \forall S_1 \neq S_2 \in \textbf{S}\}.
\end{equation}
It should be noted that the binary distance structure may have repeated elements and the order of the elements does not matter. If two binary distance structures have the same elements and the numbers of each element are the same, then they are considered as the same binary distance structure. It was found that to make $\textbf{G}^F_{K_g}\supseteq  \textbf{F}^T_{K_f}$, $\textbf{G}^F$ should have a subset with the same binary distance structure as $\textbf{F}^T$. In other words,
\begin{equation*}
 {\text {\bf Constraint 2:}} \hspace{2em} \exists \textbf{S} \subset {\textbf{G}^F},\ s.t. \ \textbf{D}_{\textbf{S}} = \textbf{D}_{\textbf{F}^T}
\end{equation*}
The proof is detailed in the appendix.

Let us use Constraint 2 to check whether right keys exist for Examples 1 and 3 in the last subsection.

\begin{enumerate}
\item In Example 1, $\textbf{F}^T$ = $\{0,1,2,3\}$ and $\textbf{G}^T$= $\{0,8,9,11,10\}$. From \eqref{Ds}, it can be computed that  $\textbf{D}_{\textbf{F}^T}=[1,2,3,3,2,1,1,2,3,3,2,1]$. A subset $\textbf{S}$ of $\textbf{G}^F$ that has the same binary structure as $\textbf{F}^T$ is $\{12, 13, 14, 15\}$. $\textbf{D}_{\textbf{S}} = [1,2,3,3,2,1,1,2,3,3,2,1]$. Hence, this block has right keys. The method to find the right keys will be presented in Section IV. It can be found that one of the right keys is $K_f = [0,0,0,0]$ and $K_g = [0,0,0,1]$
\item In Example 3, $\textbf{F}^T$ = $\{0,1,2,3\}$ and hence $\textbf{D}_{\textbf{F}^T}=[1,2,3,3,2,1,1,2,3,3,2,1]$. However, there is no subset of $\textbf{D}_{\textbf{G}^F}$ having the same binary structure as $\textbf{F}^T$. Hence right key does not exist.
\end{enumerate}

The proposed constraints for type-0 blocks can be easily extended to design type-1 G-Anti-SAT blocks. For type-1 blocks, Constraints 1 and 2 should be modified as
\begin{equation*}
    \exists F^F\in\textbf{F}^F, G^F\in \textbf{G}^F \hspace{0.15cm}s.t. \hspace{0.15cm}\textbf{D}_{F^F-\textbf{F}^F} \cap \textbf{D}_{G^F-\textbf{G}^F} = \varnothing,
\end{equation*}
\begin{equation*}
  \exists \textbf{S} \subset {\textbf{G}^T}\ s.t.\  \textbf{D}_{\textbf{S}} = \textbf{D}_{\textbf{F}^F}
\end{equation*}

\section{Generalized Anti-SAT Block Design Using K-maps}
This section proposes methods for designing the true sets for type-0 blocks that satisfy Constraint 1, 2 and finding right keys. The proposed methods are developed using K-maps. The elements in the true sets are mapped to the cells in the K-map. Accordingly, designing the true sets is translated to grouping the cells in the K-map. Whether the constraints are satisfied can be easily observed from the K-map. Also K-maps help to design blocks with lower logic complexity. The proposed design approaches can be extended similarly for type-1 blocks.

\subsection{K-map cell selection for non-complementary functions}
\begin{figure}
  \centering
  \includegraphics[width = 3.2in]{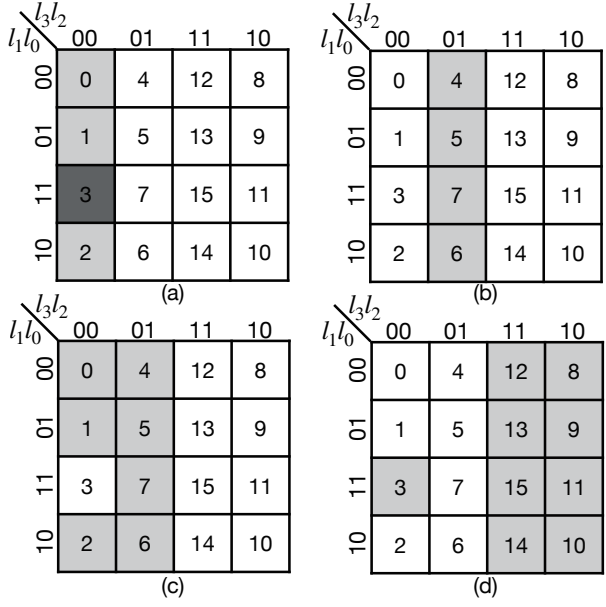}
  \caption{(a) Randomly selected cells forming $\textbf{F}^T$, {and the darker gray cell is the common cell shared with $\textbf{G}^T$}; (b) One group of cells with the same binary distance structure as those in (a); (c) Cells that should be covered by $\textbf{G}^F$ at least; (d) Cells for corresponding $\textbf{G}^T$}
  \label{rightkey-case}
\end{figure}

Let us first focus on the case that the functions $f$ and $g$ are non-complementary. Consider the G-Anti-SAT block in Fig. \ref{alternative structure} with 4-bit input $L=[l_3,l_2,l_1,l_0]$ to $f$ and $g$ as an example. The corresponding K-map has 16 cells represented as a $4\times 4$ array. $l_3l_2$ and $l_1l_0$ are used to label the columns and rows, respectively, as shown in Fig. \ref{rightkey-case}. {The minterm numbers of the cells in the K-map are also listed in the figure.}

Constraint 1 requires that there exist $F^T\in\textbf{F}^T$ and $G^T\in\textbf{G}^T$ satisfying $\textbf{D}_{F^T-\textbf{F}^T} \cap \textbf{D}_{G^T-\textbf{G}^T} = \varnothing$. When $\textbf{F}^T$ and $\textbf{G}^T$ are non-complementary, the groups of cells for $\textbf{F}^T$ and $\textbf{G}^T$ can have overlaps in the K-map and a common cell can be used as both $F^T$ and $G^T$. In a K-map, the column and row labels for each cell are distinct. Hence, adding the label of a cell to the labels of each of the other cells leads to a set of distinct labels. Accordingly, a random cell, such as the dark gray one in Fig. \ref{rightkey-case}(a), can be chosen as the cell representing both $F^T$ and $G^T$. Then $\textbf{F}^T$ and $\textbf{G}^T$ can be formed by including non-overlapping cells among the remaining cells. Also $\textbf{F}^T$ and $\textbf{G}^T$ do not need to cover all the cells.

When $f$ and $g$ are non-complementary, additional constraints need to be added to the K-map cell selection in order to satisfy Constraint 2 and hence have right keys. Let us include a quarter of the cells in the K-map for $\textbf{F}^T$. For the common cell shown in Fig. \ref{rightkey-case}(a), the cells in the leftmost column of the K-map can be randomly selected to be $\textbf{F}^T$. Although any cells including the common cell can be chosen, using adjacent cells leads to reduced logic complexity. Three groups of cells in the K-map have the same binary distance structure defined in \eqref{distance structure} as $\textbf{F}^T$. One example is the group of cells shown in Fig. \ref{rightkey-case}(b). Having $\textbf{G}^F$ include at least the cells in Fig. \ref{rightkey-case}(b) would satisfy Constraint 2. On the other hand, $\textbf{G}^T$ covers all the other cells not covered by $\textbf{G}^F$ and can only share one common cell with $\textbf{F}^T$. Therefore, $\textbf{G}^F$ also needs to cover every cell in $\textbf{F}^T$ except the common cell. For the $\textbf{F}^T$ selected in Fig. \ref{rightkey-case}(a), Fig. \ref{rightkey-case}(c) shows the $\textbf{G}^F$ satisfying these requirements. The remaining cells, as shown in Fig. \ref{rightkey-case}(d), form $\textbf{G}^T$ satisfying Constraint 1 and 2. It should be noted that there exist other choices of $\textbf{F}^T$ and $\textbf{G}^T$ satisfying Constraints 1 and 2 besides the ones that can be found by using the above method.
\begin{figure}[t]
  \centering
  \includegraphics[width=3.2in]{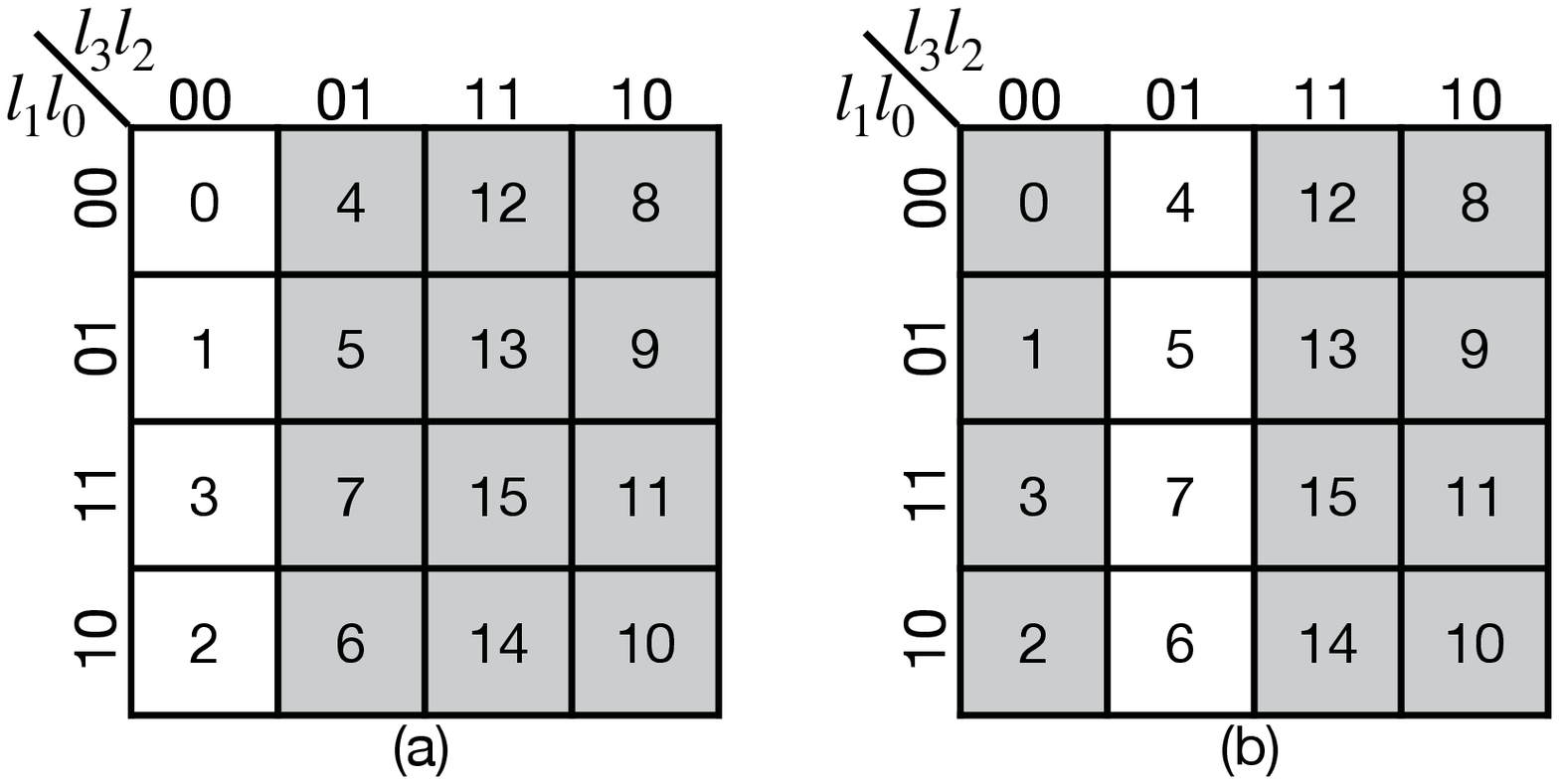}
  \caption{(a) cells of $\textbf{F}^F$; (b) cells of $\textbf{F}^F_{K_f}$ with $K_f= [0100]$}
  \label{find rightkey}
\end{figure}

The right keys $K_f$ and $K_g$ can be easily decided from the cells for $\textbf{F}^F$ and $\textbf{G}^F$ in the K-map. Constraint 2 is equivalent to \eqref{criteria3}. In the K-map, the group of cells for $\textbf{F}^F_{K_f}$ has the same shape as that for $\textbf{F}^F$, except that it is shifted and/or flipped according to the $K_f$ vector. Similarly, the group of cells for $\textbf{G}^F_{K_g}$ is that for $\textbf{G}^F$ shifted and/or flipped according to $K_g$. Then \eqref{criteria3} is translated to that the shifted and/or flipped groups for $\textbf{F}^F$ and $\textbf{G}^F$ need to cover every cell in the K-map. The vectors leading to such shifting/flipping are the right keys $K_f$ and $K_g$. For the $\textbf{F}^T$ in Fig. \ref{rightkey-case}(a), Fig. \ref{find rightkey}(a) shows the cells for the corresponding $\textbf{F}^F$. The cells for $\textbf{G}^F$ are illustrated in Fig. \ref{rightkey-case}(c). It can be seen that the union of such $\textbf{G}^F$ and $\textbf{F}^F$ covers every cell in the K-map except the one with $[l_3,l_2,l_1,l_0]=[0,0,1,1]$, which is the common cell. One way to cover every cell in the K-map is to keep the cells for $\textbf{G}^F$ unchanged, which means $K_g=[0,0,0,0]$, and use $K_f= [0,1,0,0]$, which leads to the group of cells of $\textbf{F}^F_{K_f}$ shown in Fig. \ref{find rightkey}(b). The gray cells in Fig. \ref{rightkey-case}(c) and Fig. \ref{find rightkey}(b) are $\textbf{G}^F_{K_g}$ with $K_g=[0,0,0,0]$ and $\textbf{F}^F_{K_f}$ with $K_f= [0,1,0,0]$, respectively. They cover all cells in the K-map. There are many choices of  $K_f$ and $K_g$ that satisfy \eqref{criteria3}. Another example is $K_g=[0,1,0,0]$ and $K_f= [0,0,0,0]$. It corresponds to that the $\textbf{F}^F$ in Fig. \ref{find rightkey}(a) is unchanged and {the column labels of the cells in $\textbf{G}^F$ are XORed with the 2-bit vector [0,1] to form $\textbf{G}^F_{K_g}$. Hence, in K-map, the cells for $\textbf{G}^F_{K_g}$ are those in Fig. \ref{rightkey-case}(c) flipped horizontally.}

{
Our proposed design can be easily generalized to $n$-bit input $L = [l_{n-1},l_{n-2},\cdots,l_1,l_0]$. {The cardinality of $\textbf{F}^T$ is chosen to be $2^{n-t}$} ($2\leq t\leq n-1$) to simplify the logic. To facilitate the design, the K-map can be drawn as an array of $2^{n-t} \times 2^t$ cells. Use $l_{n-t-1},\cdots l_1,l_0$ to label the rows and $ l_{n-1},\cdots l_{n-t+1},l_{n-t}$ to label the columns. Then $\textbf{F}^T$ can cover a column of the K-map, and the $f$ function is the AND operation among the $t$ bits of the column label. To satisfy Constraint 1, $\textbf{G}^T$ should include exactly one element from $\textbf{F}^T$. Besides, to satisfy Constraint 2, $\textbf{G}^T$ can include the remaining columns except one whose column label is different in one single bit from the column label of $\textbf{F}^T$ as in the example presented in Fig. \ref{rightkey-case}(d). Accordingly, $\textbf G^T$ can include $2^t-2$ columns plus one cell in the K-map and $|\textbf{G}^T| = 2^n-2^{n-t+1}+1$. Without loss of generality, let $\textbf{F}^T$ cover the column with all `0' label. Then
\begin{equation}\label{fL}
f(L) = \bar l_{n-t}\&\bar l_{n-t+1}\&\cdots\&\bar l_{n-1}.
\end{equation}
Assume that the two columns not in $\textbf G^T$ are different in bit $l_q$ in their labels. Then the logic expression for the part of $\textbf G^T$ that consists of the $2^t-2$ columns is
\begin{equation}\label{g1L}
g_1(L) =l_{n-t}+\cdots+l_{q-1}+l_{q+1}+\cdots+l_{n-1}.
\end{equation}
The one cell from $\textbf{F}^T$ that is also included in $\textbf{G}^T$ can be grouped with the other cells in $\textbf{G}^T$ to simplify the logic expression. If the cell whose row label is all `0' is picked as this common cell, then the logic expression covering this cell is
\begin{equation}\label{g2L}
g_2(L) = \bar l_q \& \bar l_0\& \bar l_1\& \cdots \& \bar l_{n-t-1}.
\end{equation}
Overall, $g(L)=g_1(L)+g_2(L)$. Of course, a different column can be chosen for $\textbf F^T$ and an alternative common cell can be used. In this case, the literals in the $f(L)$ and $g(L)$ functions need to be complemented accordingly. In order to increase the corruptibility of the output signal, the product of $|\textbf{F}^T|$ and $|\textbf{G}^T|$ needs to be as large as possible. The best design for achieving this goal for the $n$-bit non-complementary design we found so far is to let $|\textbf{F}^T| = 2^{n-2}$, $|\textbf{G}^T| = 2^{n-1}+1$. }

{
The right keys for the above non-complimentary design can be also obtained by flipping/shifting the false  sets of $f$ or $g$ to make their union cover all the cells in the $n$-bit K-map. Equivalently, this means that there should be no overlap between $\textbf{G}^T_{K_g}$ and $\textbf{F}^T_{K_f}$. To meet this requirement, the right key can be any $K_f = [K_{f,n-1}, K_{f,n-2},\cdots,K_{f,0}]$ and $K_g = [K_{g,n-1}, K_{g,n-2}, \cdots, K_{g,0}]$ such that $K_{f,i}=K_{g,i}$ ($n-t\leq i\leq n-1, i \neq q$) and $K_{f,q} = \overline{K_{g,q}}$. The bits in the other positions can be either `0' or `1'.}

{
In summary, for a chosen cardinality $|\textbf F^T|=2^{n-t}$ ($2\leq t\leq n-1$), an $n$-bit non-complementary G-Anti-SAT block can be designed according to the following steps.
\begin{enumerate}
\item Select a column in the $2^{n-t} \times 2^t$ K-map for $\textbf F^T$; Select one cell in this column as the common cell; Select an integer $q$ in the range of $[n-t,n-1]$;
\item Design $f(L)$ and $g(L)=g_1(L)+g_2(L)$ using \eqref{fL}, \eqref{g1L} and \eqref{g2L}. If a bit in the labels for the column and common cell selected in Step 1) is `0', then use the literals as in \eqref{fL}, \eqref{g1L} and \eqref{g2L}. If the bit is `1', complement the corresponding literals in the equations.
\item Pick any $K_f$ and $K_g$ such that they are different in bit $q$ and are the same in bits $n-t$ through $n-1$ to be a right key.
\end{enumerate}
}

{
It should be noted that $\textbf F^T$ and $\textbf G^T$ do not have to cover entire columns as in the design process explained above. This would provide more variation on the design at the cost of higher logic complexity.}

\subsection{K-map cell selection for complementary functions}
\begin{figure}
  \centering
  \includegraphics[width=3.2in]{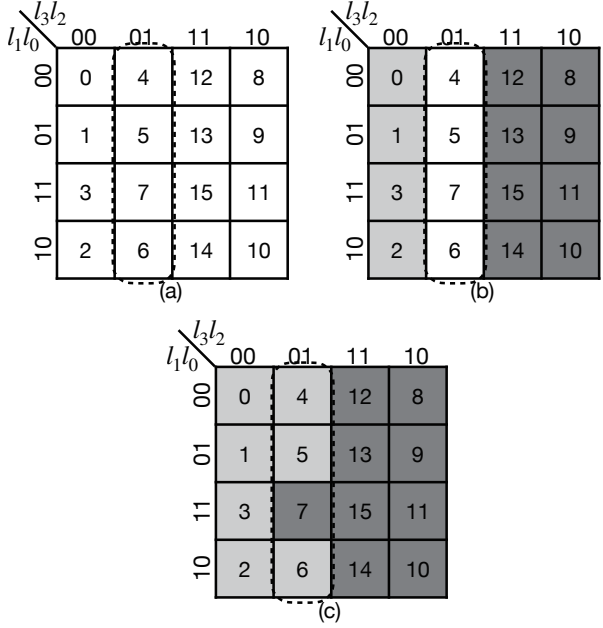}
  \caption{(a) The column circled in dashed line is the dividing column; (b) Splitting the other columns between $\textbf{F}^T$ (light gray) and $\textbf{G}^T$ (dark gray); (c) Splitting the cells in the dividing column between $\textbf{F}^T$ and $\textbf{G}^T$ }
  \label{complementary}
\end{figure}

When $f$ and $g$ are complementary, $\textbf{F}^T$ and $\textbf{G}^T$ should not have any common cells and should cover all the cells in the K-map. First pick a random column as shown in Fig. \ref{complementary}(a). This column is referred to as the dividing column in this paper. The labels for each column in the K-map are distinct. Hence, adding the column label of the dividing column to the labels of the other columns results in a set of distinct nonzero vectors. This means that splitting the other columns between $\textbf{F}^T$, $\textbf{G}^T$ and picking $F^T$, $G^T$ from the dividing column would satisfy Constraint 1. For example, in Fig. \ref{complementary}(b), the first column is put in $\textbf{F}^T$ and the third and fourth columns are put in $\textbf{G}^T$.

Next, the cells in the dividing column should be split between $\textbf{F}^T$ and $\textbf{G}^T$. Put one cell of this column in one set and the others in the other set. Without loss of generality, the one cell is put in $\textbf{G}^T$ and the other cells are put in $\textbf{F}^T$, as shown by the example in Fig. \ref{complementary}(c). The cell of  $\textbf{G}^T$ in the dividing column can be used as $G^T$ and any other cells in the dividing column can be used as $F^T$. {The column labels of any two cells in the same column are the same.} Hence, adding $F^T$ to any other cells in the dividing column would result in a zero column label, and any vector in $\textbf{D}_{G^T-\textbf{G}^T}$ is different from those in $\textbf{D}_{F^T-\textbf{F}^T}$. As a result, {splitting $\textbf{F}^T$ and $\textbf{G}^T$ in this way satisfies Constraint 1.}

As mentioned previously, when $f$ and $g$ are complementary, any $K_f=K_g$ can be used as a right key. Constraint 2 does not need to be considered in this case.

{The complementary G-Anti-SAT design can be also easily generalized to $n$-bit input $L = [l_{n-1},l_{n-2},\cdots,l_1,l_0]$. The cardinality of $\textbf{F}^T$ is chosen to be in the format of $2^{n-t}-1\ (1\leq t \leq n-1)$. Then $|\textbf{G}^T| = 2^n-2^{n-t}+1$. To facilitate the design, the K-map is drawn as an array of $2^{n-t}\times 2^t$ cells and the columns and rows are labeled in the same way as in the non-complementary design. As in the example of Fig. \ref{complementary}(a), a column is first selected as the dividing column. To satisfy Constraint 1, one cell in the dividing column and the remaining columns are put in $\textbf{G}^T$, and the remaining cells in the dividing column are included in $\textbf{F}^T$. Without loss of generality, choose the column with all `0' label as the dividing column. Then the logic expression for the part of $\textbf{G}^T$ that consists of the remaining $2^t-1$ columns is
\begin{equation}\label{comp g1}
    g_1(L) = l_{n-t}+l_{n-t+1}+\cdots+l_{n-1}.
\end{equation}
The cell in the dividing column that is included in $\textbf{G}^T$ can be grouped with the other cells in the same row of the K-map that are also in $\textbf{G}^T$. In the case that this cell has all `0' in its label, the logical expression covering this cell is
\begin{equation}\label{comp g2}
    g_2(L) = \bar l_0\& \bar l_1\& \cdots \& \bar l_{n-t-1}
\end{equation}
Overall, $g(L) = g_1(L)+g_2(L)$, $f(L) = \overline {g(L)}$. A different column can be chosen as the dividing column and an alternative cell can be put in $\textbf{G}^T$. In this case, the literals in $g_1(L)$ and $g_2(L)$ formulas need to be complemented accordingly. For the $n$-bit complementary design, the maximum corruptibility in the output is achieved when $|\textbf{F}^T|=2^{n-1}-1$ and $|\textbf{G}^T|=2^{n-1}+1$.
}

{
In summary, for a chosen cardinality $|\textbf{F}^T|=2^{n-t}-1$ ($1\leq t\leq n-1$), an $n$-bit complementary G-Anti-SAT block can be designed according to the following steps.
\begin{enumerate}
    \item Select a column of the $2^{n-t}\times 2^t$ K-map as the dividing column; Select one cell in this column to put in $\textbf{G}^T$
    \item Design $g(L) = g_1(L)+g_2(L)$ using \eqref{comp g1} and \eqref{comp g2}. Then $f(L) = \overline{g(L)}$. If a bit in the labels for the dividing column and the cell selected in Step 1) is `0', then use the literals as in \eqref{comp g1} and \eqref{comp g2}. If the bit is `1', complement the corresponding literals in the equations.
    \item Any $K_f=K_g$ can be used as the right key.
\end{enumerate}
}

{
Similarly, $\textbf F^T$ and $\textbf G^T$ do not have to cover entire columns as in the design process explained above. Other designs satisfying Constraint 1 are possible, although the logic complexity may be higher}

\section{Experiments, Analysis, and Comparisons}
This section evaluates the effectiveness of our proposed G-Anti-SAT scheme against the SAT, AppSAT, and removal attacks. The SAT attack tool in \cite{SAT} based on the Lingeling SAT solver is used in our experiments to test the running time of the SAT attack. The CPU time is limited to 10 hours, and the experiments are run over an Intel Core i7 with 4GB RAM. {{The resiliency of our logic-locking blocks against the AppSAT attack is evaluated by analyzing the average corruptibility and the corruptibility profile of the approximate keys that can be returned by the attack.} This analysis is enabled by the tool in \cite{Florida} that can return an approximate key after any given number of SAT attack iterations.} The evaluation for the removal attack resistance is done by calculating the ADS values as in \cite{Removal}. At the end, the area requirement of our design from synthesis reports is compared to that of prior designs.

\subsection{SAT attack resistance analysis}
\begin{table}[t]
  \centering
  \caption{Number of iterations and time needed by the SAT attack to decrypt the G-Anti-SAT and Anti-SAT blocks}
  \label{SAT vs m1 m2}
  \begin{tabular}{c|c|c|c|c}
  \hline
\quad & & $n=8$ & $n=12$ & $n=16$\\
  \hline
  \multirow{2}{2.6cm}{non-complementary G-Anti-SAT} & \# of iterations  & 255 & 4095 & -\\
  \cline{2-5}
  & time (second) & 0.44 & 66.21 & timeout\\
  \hline
  \multirow{2}{2.6cm}{complementary G-Anti-SAT} & \# of iterations  & 255 & 4095 & -\\
  \cline{2-5}
  & time (second) & 0.80 & 166.42 & timeout\\
  \hline
  \multirow{2}{2.6cm}{Anti-SAT block  \cite{XieMitigating}} & \# of iterations  & 255 & 4095 & -\\
  \cline{2-5}
  & time (second)& 0.82 & 175.74 & timeout\\
  \hline
  \end{tabular}
\end{table}

The SAT attack is applied to decrypt the proposed G-Anti-SAT blocks, and the number of iterations and time are listed in Table \ref{SAT vs m1 m2} for the designs with different numbers of input bits. For comparison, the Anti-SAT block is also simulated in the same hardware environment and the results are included in Table \ref{SAT vs m1 m2}. It can be observed from this table that our design achieves the same resistance to the SAT attack in terms of the number of iterations, which matches our previous analysis. The reason that the time consumed by the attack on the non-complementary G-Anti-SAT block is shorter than those on the complementary G-Anti-SAT and Anti-SAT blocks is because that the non-complementary design has less complicated logic. As a result, the complexity to construct and solve the corresponding CNF formula is lower. From the table, we can see that for 16-bit input, the time needed by the SAT attack to decipher the G-Anti-SAT block is already over 10 hours. Also the time needed for the attack increases very fast with the input size.

\subsection{AppSAT attack resistance analysis}
\begin{figure}
    \centering
    \includegraphics[width=3.5in]{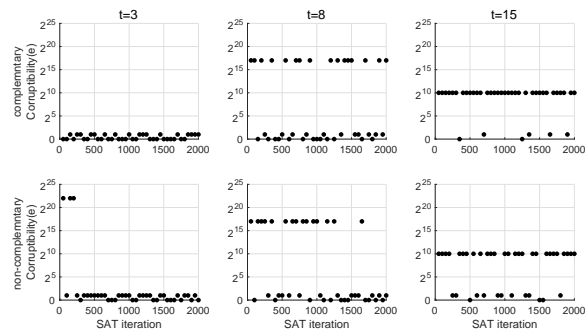}
    \caption {Corruptibility of the returned keys over SAT attack iterations for G-Anti-SAT blocks with $n=25$}
    \label{AppSAT resistance}
\end{figure}

\begin{table}
  \renewcommand\arraystretch{1.3}
  \centering
  \caption{Corruptibilities ($e$) of wrong keys and average corruptibility for $n$-bit G-Anti-SAT designs}
  \label{comp corruptibility}
  \begin{tabular}{@{}c|c@{}|c@{}|c@{}}
    \hline
    \quad & \textbf{\# of wrong keys} & \textbf{Corruptibility}& \textbf{Average} \\
& & ($e$)& \textbf{corruptibility} \\
    \hline
    Complementary & $2^{2n}-2^{2n-t}$ &  $2^{n-t}-1$ &  $2^{n-t}-2^{n-2t}$ \\
    \cline{2-3}
    G-Anti-SAT&  $2^{2n-t}-2^n$ & 1  & \\\hline

    Non-complementary &$2^{2n}-2^{2n-t+1}$ & $2^{n-t}$ &  $2^{n-t}-2^{n-2t+1}$ \\
    \cline{2-3}
    G-Anti-SAT&  $2^{2n-t}$ & 1 & \\\hline
  \end{tabular}
\end{table}

The corruptibility of each wrong key, which is the number of input patterns leading to the wrong output, is 1 in the Anti-SAT scheme \cite{XieMitigating}. Adopting the relaxation on the wrong key sets, our proposed G-Anti-SAT scheme can increase the corruptibility of a large portion of the wrong keys. {Hence, in each of the two proposed schemes, the wrong keys have two possible corruptibilities. The corruptibilities denoted by $e$ and the number of wrong keys with those corruptibilities are listed in Table \ref{comp corruptibility} for each of the proposed G-Anti-SAT schemes with $n$-bit input and parameter $t$.} The average corruptibility is computed as the sum of the corruptibility of each wrong key divided by the total number of wrong keys. The average corruptibility is maximized in the complementary design when $t=1$, and it is around $2^{n-2}$.

The average corruptibility is not the single criterion to evaluate the resiliency to the AppSAT attack. {It is possible that a large average corruptibility is resulted from a small portion of wrong keys with very high corruptibility. In this case, those high-corruptibility keys will be excluded by the AppSAT attack in a few iterations, and the key returned by the AppSAT attack will have low corruptibility. As listed in Table \ref{comp corruptibility}, each wrong key in our G-Anti-SAT design can have one of the two possible corruptibilities, which are referred to as the high corruptibility and low corruptibility in the remainder of the discussion. In our design, a larger $t$ leads to lower average corruptibility but a larger portion of wrong keys with high corruptibility. As a result, when the AppSAT attack is applied, the chance of returning a high-corruptibility key is larger, although the high corruptibility for the design with $t$ is smaller than that of the design with $t'<t$.}

{
The running time of the AppSAT attack depends on many parameters, such as the number of random patterns for each round, error threshold, and settlement count \cite{AppSAT}. Besides, these parameters can be adjusted according to the logic-locking scheme. Hence, the running time of the AppSAT attack is not a good measurement of the resiliency towards this attack. Instead, the corruptibility of the key returned by the AppSAT attack affects the usability of the key a lot. Hence, in the following, the focus is given to the corruptibilities of the keys that can be returned after different numbers of SAT iterations.
}

{
The corruptibility profiles of 25-bit-input G-Anti-SAT designs with different $t$ are presented in Fig. \ref{AppSAT resistance}. In these plots, the $x$-axis is the iteration number in the SAT attack. A key is returned after every 50 iterations. The $y$-axis is the corruptibility of the returned key. For $t=3$, the high corruptibility is very large, which is $2^{25-3}=2^{22}$ and $2^{22}-1$ for the non-complementary and complementary designs, respectively, from Table \ref{comp corruptibility}. However, the portion of the wrong keys with this high corruptibility is very small. As a result, as shown by the two plots in the first row of Fig. \ref{AppSAT resistance}, the chance of returning a low-corruptibility key is very high. As $t$ increases to 8 and 15, the high corruptibility is reduced to around $2^{17}$ and $2^{10}$, respectively. Nevertheless, the portion of the wrong keys with high corruptibility increases exponentially with $t$. Hence, as it can be observed from the plots in the second and third {columns} of Fig. \ref{AppSAT resistance}, the chance of returning a high-corruptibility key becomes much higher for larger $t$. It should be noted that the value of the high corruptibility decreases for larger $t$. When $t=n-1$, the value of the high corruptibility reduces to 1. Hence using a large $t$ close to $n-1$ does not lead to better resiliency to the AppSAT attack, and a medium $t$ strikes a balance between the value of the corruptibility and the chance of returning a wrong key with good corruptibility.
}

\begin{figure}[t]
    \centering
    \includegraphics[width=3.5in]{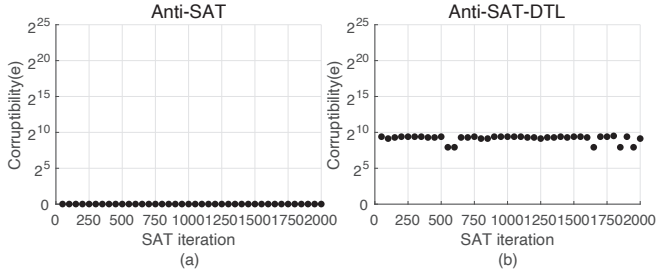}
    \caption{Corruptibilities of the returned keys over SAT attack iterations for logic-locking blocks with $n=25$. (a) Anti-SAT; (b) Anti-SAT-DTL with 6 OR gates in the first layer}
    \label{Anti-SAT-DTL}
\end{figure}

\begin{table*}
  \renewcommand\arraystretch{1.3}
  \centering
  \caption{Corruptibility for $n$-bit Anti-SAT-DTL scheme}
  \label{DTL}
  \begin{tabular}{c|c|c}
    \hline
    \textbf{Gate-type} & \textbf{Corruptibility ($e$)} &\textbf{Average corruptibility} \\
    \hline
    XOR & $[{(2(2^{2^{l-1}}-1))}^{r-1} ,{(2(2^{2^{l-1}}-1))}^{r}]$ & $(2(2^{2^{l-1}}-1))^r(2^n-(2(2^{2^{l-1}}-1))^r)/2^{n}$ \\
    \hline
    OR/NAND & $[{(1+2(2^{2^{l-1}}-1))}^{r-1}, {(1+2(2^{2^{l-1}}-1))}^{r}]$ & $(1+2(2^{2^{l-1}}-1))^r(2^n-(1+2(2^{2^{l-1}}-1))^r)/2^{n}$ \\
    \hline
  \end{tabular}
\end{table*}

{
For reference, the corruptibility profiles of the Anti-SAT and Anti-SAT-DTL scheme \cite{AppSAT} with $n=25$-bit input are shown in Fig. \ref {Anti-SAT-DTL}. As expected, for the Anti-SAT design, the corruptibility of the wrong key returned is always 1. In the Anti-SAT-DTL scheme, the AND/NAND trees in the Anti-SAT block are replaced by $n$-bit DTL. When $r$ AND gates in layer $l$ ($1\leq l \leq \lceil log_2  n \rceil$) are replaced by XOR/OR/NAND gates, it has been derived in \cite{AppSAT} that the corruptibilities of the wrong keys are in the range shown in Table \ref{DTL}. The average corruptibility can be also calculated as listed in this table. Fig. \ref{Anti-SAT-DTL}(b) is for the case that 6 AND gates in the first layer are replaced by OR gates. Since the corruptibilities, $e$, of all wrong keys are high in the Anti-SAT-DTL design, it is more resilient to the AppSAT attack compared to our G-Anti-SAT scheme. However, its better AppSAT attack resiliency comes at the cost of reducing the number of iterations needed by the SAT attack to $2^n/e$ \cite{AppSAT}.
}

\subsection{Removal attack resistance analysis}
{
The SPS removal attack utilizes the ADS values. The ADS value of a gate is in the range of [0,1). However, it is rare that a circuit contains a gate whose ADS value is very close to 1 as the last gate of the Anti-SAT block. Hence, the last gate can be identified by sorting the ADS values. The ADS values of most gates in a circuit are in a medium range. By tuning the parameter $t$, the cardinalities of the true sets of the $f$ and $g$ functions of our G-Anti-SAT design change. Accordingly, the ADS value of the last gate varies and it is well-hidden among the ADS values of the other gates. As a result, unlike the Anti-SAT design, it is difficult to identify the last gate in our G-Anti-SAT design by finding the gate with either the largest or smallest ADS value. For example, a non-complimentary G-Anti-SAT block can be designed to have $2^{n-2}$ and $2^{n-1}+1$ as the cardinalities of $\textbf F^T$ and $\textbf G^T$, respectively. In this case, the ADS value of the last gate is $|\frac{2^{n-2}}{2^n} - \frac{2^{n-1}+1}{2^n}| \approx 0.25$.
}

In order to resist the SPS attack, structural and functional obfuscations can be inserted into the Anti-SAT block \cite{XieMitigating}. However, the keys for structural and functional obfuscation can be separated from those for the Anti-SAT block by the AGR attack \cite{Removal}. After that, the last gate and the correct output signal can still be obtained. The reason that the keys can be separated is because that the keys to the Anti-SAT scheme has low corruptibility while those for the obfuscation schemes have high corruptibility. In our G-Anti-SAT designs, the output already has higher corruptibility even if a larger $t$ is used to achieve better AppSAT resistance. Hence, if our design is combined with structural and functional obfuscations, the keys are harder to be separated by using the AGR attack and even better resistance to the removal attack can be achieved.

\subsection{Complexity comparison}
\begin{table}[h]
\centering
\caption{Areas of logic-locking blocks with $t=3$ synthesized using TSMC 65$nm$ process with $4ns$ timing constraint}
\begin{tabular}{c|c|c|c}
\hline
& $n=8$ & $n=16$ & $n=25$ \\\hline\hline
non-complementary & 91.440  & 153.000 & 222.480\\
G-Anti-SAT(${\mu m}^2$)&  &  & \\
\hline
complementary & 131.400  & 254.880 & 393.480\\
G-Anti-SAT(${\mu m}^2$)&  &  & \\
\hline
Anti-SAT (${\mu m}^2$) & 129.600 & 253.080 & 393.120\\
\hline
\end{tabular}
\label{area}
\end{table}

{Our proposed complementary and non-complementary G-Anti-SAT designs with different $n$ and $t=3$ are synthesized using TSMC 65$nm$ CMOS process under 4$ns$ timing constraints, and the results are listed in Table \ref{area}. For comparison, the area of the Anti-SAT design is also included in this table. Our complementary design has very similar area as the Anti-SAT design. On the other hand, our non-complementary design is much smaller. The reason is that the true set of one function in the non-complementary design consists of a whole column of the K-map and hence its logic expression is substantially simpler. To show how the complexity of our designs change with $t$, synthesis results for $n=25$ and different $t$ are listed in Table \ref{area_noncomp}. It can be observed that, when larger $t$ is adopted, the area for the non-complementary case increases. This is because more literals are included in the $f$ function as shown in \eqref{fL}. On the other hand, the area requirement for the complementary case remains similar for different $t$ since changing $t$ only leads to switching literals between the $g_1(L)$ and $g_2(L)$ functions in \eqref{comp g1} and \eqref{comp g2}.}

\begin{table}[h]
\centering
\caption{Areas of G-Anti-SAT blocks with $n=25$ synthesized using TSMC 65$nm$ process with $4ns$ timing constraint}
\begin{tabular}{c|c|c|c}
\hline
& $t=3$ & $t=8$ & $t=15$ \\\hline\hline
non-complementary & 222.480  & 261.000 & 316.800\\
G-Anti-SAT(${\mu m}^2$)&  &  & \\
\hline
complementary & 393.480  & 394.200 & 392.760\\
G-Anti-SAT(${\mu m}^2$)&  &  & \\
\hline
\end{tabular}
\label{area_noncomp}
\end{table}

\section{Discussions}
Our proposed G-Anti-SAT designs enjoy great flexibility on the $f$ and $g$ functions. They do not have to be AND/NAND tree or complementary to make the query count needed in the SAT attack exponential. Unlike previous designs, the true sets of the functions can be chosen to have larger cardinalities and hence increase the corruptibility without sacrificing the SAT-attack resilience.


There is another attack called the bypass attack \cite{Bypass}. The main idea is to add a bypass circuit that inverts the wrong output and nullifies the effect of the wrong key on the encrypted circuit. {The complexity of the bypass circuit increases linearly with $N_p$, the number of input patterns that generate the wrong output. In the Anti-SAT design, each wrong key only leads to wrong output for one and only one input pattern and hence $N_p=1$. However, for an $n$-bit complementary G-Anti-SAT design with $|\textbf{G}^T| = 2^{n-t}-1$, there are $2^{2n}-2^{2n-t}$ wrong keys with $2^{n-t}-1$ as the corruptibility from Section V.B. If any of those wrong keys is used, $N_p=2^{n-t}-1$ and hence the overhead of the corresponding bypass circuit is much higher. Similarly, to apply the bypass attack on the non-complementary G-Anti-SAT design, the area overhead would be also very high due to the high corruptibility of the wrong keys.}

{It was also realized that the wrong key sets may have overlaps in the recent CAS-Lock design\cite{CAS}. However, unlike our G-Anti-SAT design, the CAS-Lock design uses specific complementary functions that are implemented by a cascade of AND and OR gates. Our design is more generalized in the sense that i) both complementary and non-complementary functions are allowed, and ii) the functions have a very large number of variations and they are not limited to certain type of structures. Besides the design procedures presented in Section IV, other $f$ and $g$ functions are possible by choosing true sets of other cardinalities and/or corresponding to alternative groups of cells in the K-map. Actually, the CAS-Lock design is a special case of our complementary design with the true set of $f$ consisting of consecutive cells of the K-map in numerical order starting with the all-`0' cell. The CAS-Lock block can be attacked by the CAS-Unlock attack\cite{CAS-unlock}, which uses either all '0' or all `1' as the key. Such a key leads to the correct output for all input patterns because the right keys for the two complementary functions are the same. This attack applies to the Anti-SAT and our complementary designs as well. To address this issue, it was proposed in \cite{DCAS-unlock} to replace some of the XOR gates at the inputs of the function blocks randomly by XNOR gates \cite{DCAS-unlock}. This scheme can be also adopted for our complementary G-Anti-SAT design. Nevertheless, our non-complementary design is not subject to the CAS-Unlock attack since the keys to the two functions are not the same. Another advantage of our design is that no specific structure is required in the two functions. As a result, our design is immune to any attacks that try to utilize specifics of the functions or structures.}

\section{Conclusions}
In this paper, novel G-Anti-SAT schemes have been proposed by relaxing the constraints on the wrong key sets. Our schemes allow great flexibility on the two function blocks, which can be also non-complementary. Our designs are always resilient to the SAT attack. Moreover, by choosing functions whose true sets have larger cardinalities, the output corruptibility is effectively increased. As a result, our design has better resistance to the AppSAT and other attacks without compromising the SAT-attack resiliency. Methodologies have also been provided for designing the G-Anti-SAT blocks and deciding the right keys using the K-map. Future work will monitor new attacks and extend our proposed designs.

\appendices
\section{}
The true sets of non-complementary $f$ and $g$ have a common cell. Hence, $\textbf{F}^F \cup \textbf{G}^F \subset \textbf{X}_n$. Accordingly, $\textbf{F}^T_{K_f}$ needs to be a subset of $\textbf{G}^F_{K_g}$ in order to satisfy \eqref{criteria3}. Let $\textbf{C}$ be a subset of $\textbf{G}^F_{K_g}$ that equals $\textbf{F}^T_{K_f}$. Assume that $|\textbf{C}| =|\textbf{F}^T_{K_f}|=m$, and for elements $C_i\in \textbf{C}$ and $F_{K_f,i}^T\in \textbf{F}^T_{K_f}$ ($i=1,2,\cdots,m$)
\begin{align*}
  C_1 &= {F_{K_f,1}^T}\\
  &\vdots \\
  C_n &= {F_{K_f,m}^T}.
\end{align*}
Since $\textbf{C} \subset \textbf{G}^F_{K_g}$, according to the definition of $\textbf{G}^F_{K_g}$, $C_i$ can be written as ${G^F_i} \oplus K_g$ for some ${G^F_i}\in \textbf G^F$. Similarly, $ F_{K_f,i}^T={F^T_i} \oplus K_f$. Then the above equations can be rewritten as
\begin{align*}
  {G^F_1} \oplus K_g &= {F^T_1} \oplus K_f \\
  &\vdots \\
  {G^F_m} \oplus K_g &= {F^T_m} \oplus K_f.
\end{align*}
Moving $K_g$ from the left side to the right side of the equations, it can be derived that
\begin{equation}\label{append1}
\begin{aligned}
  {G^F_1} &= {F^T_1} \oplus K\\
  &\vdots \\
  {G^F_m} &= {F^T_m} \oplus K ,
\end{aligned}
\end{equation}
where $K = K_f \oplus K_g$. Adding any two equations listed in \eqref{append1} leads to ${G^F_i} \oplus {G^F_j} = {F^T_i} \oplus {F^T_j}$. Let $\textbf{S}=\{G^F_1, G^F_2, \cdots, G^F_m\}$. Apparently, $\textbf{S}\subset \textbf{G}^F$. Therefore,
\begin{equation}\label{key}
\exists \textbf{S}\subset \textbf{G}^F,\ s.t.\  \forall S_i, S_j,\ S_i \oplus S_j = {F^T_i} \oplus {F^T_j}.
\end{equation}

According to the definition of binary distance structure in \eqref{Ds}, \eqref{key} can be translated to Constraint 2.

%


\begin{thebibliography}{2}

\bibitem{piracy}
M. Rostami, F. Koushanfar and R. Karri, ``A primer on hardware security: models, methods, and metrics,'' {\it Proc. of the IEEE}, vol. 102, no. 8, pp. 1283-1295, Aug. 2014.

\bibitem{counterfeiting}
U. Guin, {\it et. al.}, ``Counterfeit integrated circuits: a rising threat in the global semiconductor supply chain,'' {\it Proc. of the IEEE}, vol. 102, no. 8, pp. 1207-1228, Aug. 2014.

\bibitem{Cocchi}
R. P. Cocchi, J. P. Baukus, L. W. Chow and B. J. Wang ``Circuit camouflage integration for hardware IP protection," {\it Proc. of ACM/EDAC/IEEE Design Automation Conf.}, pp. 1-5, San Francisco, CA, U.S.A., 2014.

\bibitem{Camouflaging1}
J. Rajendran, M. Sam, O. Sinanoglu and R. Karri, ``Security analysis of integrated circuit camouflaging,'' {\it Proc. ACM SIGSAC Conf. on Computer \& Commun. Security}, pp. 709-720, New York, U.S.A, 2013.

\bibitem{reverse-engineering}
R. Torrance and D. James, ``The state-of-the-art in semiconductor reverse engineering,'' {\it Proc. of ACM/EDAC/IEEE Design Automation Conf.}, pp. 333-338, New York, NY, U.S.A., 2011.

\bibitem{RoyEPIC}
J. A. Roy, F. Koushanfar, and I. L. Markov, ``EPIC: Ending piracy of integrated circuits,'' {\it Proc. Conf. on Design, Automation and Test in Europe}, pp. 1069-1074, Munich, Germany, 2008.


\bibitem{Fault analysis}
J. Rajendran {\it et. al.}, ``Fault analysis-based logic encryption,'' {\it IEEE Trans. on Computers,} vol. 64, no. 2, pp. 410-424, Feb. 2015.

\bibitem{Yasin-logic}
M. Yasin, J. Rajendran, O. Sinanoglu and R. Karri ``On improving the security of logic locking,'' {\it IEEE Trans. on Computer-Aided Design of Integrated Circuits and Syst.}, vol. 35, no. 9, pp. 1411-1424, Sept. 2016.

\bibitem{RajendranSecurity}
J. Rajendran, Y. Pino, O. Sinanoglu and R. Karri ``Security analysis of logic obfuscation,'' {\it Proc. of ACM/EDAC/IEEE Design Automation Conf.}, pp. 83-89, San Francisco, CA, U.S.A, 2012.

\bibitem{PUF}
J. B. Wendt and M. Potkonjak, ``Hardware obfuscation using PUF-based logic,'' {\it IEEE/ACM Intl. Conf. on Computer-Aided Design}, pp. 270-271, San Jose, CA, U.S.A., 2014.


\bibitem{Lee-logic}
Y. Lee and N. A. Touba, ``Improving logic obfuscation via logic cone analysis,'' {\it Latin-American Test Symposium,} pp. 1-6, Puerto Vallarta, Mexico, 2015.

\bibitem{Entanglement}
S. Khaleghi, K. Zhao and W. Rao, ``IC piracy prevention via design withholding and entanglement,'' {\it Asia and South Pacific Design Auto. Conf.}, pp. 821-826, Chiba, Japan, 2015.

\bibitem{LUT2}
B. Liu and B. Wang, ``Embedded reconfigurable logic for ASIC design obfuscation against supply chain attacks,'' {\it Proc. Conf. on Design, Automation and Test in Europe}, pp. 1-6, Dresden, Germany, 2014.

\bibitem{SAT}
P. Subramanyan, S. Ray, and S. Malik, ``Evaluating the security of logic encryption algorithms,'' {\it Proc. IEEE Intl. Symp. on Hardware Oriented Security and Trust}, pp. 137-143, Washington DC, U.S.A., 2015.

\bibitem{XieMitigating}
Y. Xie and A. Srivastava, ``Anti-SAT: mitigating SAT attack on logic locking,'' {\it  IEEE Trans. on Computer-Aided Design of Integrated Circuits and Syst.}, vol. 38, no. 2, pp. 199-207, Feb. 2019.

\bibitem{YasinSarlock}
M. Yasin, B. Mazumdar, J. Rajendran and O. Sinanoglu ``SARlock: SAT attack resistant logic locking,'' {\it Proc. IEEE Intl. Symp. on Hardware Oriented Security and Trust}, pp. 236-241, McLean, VA, U.S.A., 2016.

\bibitem{AppSAT}
K. Shamsi, {\it et. al.}, ``On the Approximation Resiliency of Logic Locking
and IC Camouflaging Schemes,'' {\it  IEEE Transactions on Information Forensics and Security,} vol.14, no. 2, pp. 347-359, Feb. 2019

\bibitem{Removal}
M. Yasin, B. Mazumdar, O. Sinanoglu and J. Rajendran ``Removal attacks on logic locking and camouflaging techniques,'' {\it IEEE Trans. on Emerging Topics in Computing}, pp. 1-1.

\bibitem{bit-flip}
Y. Shen, A. Rezaei and H. Zhou, ``SAT-based bit-flipping attack on logic encryptions,''{\it Proc. Conf. on Design, Automation and Test in Europe}, pp. 629 -632, Dresden, Germany, 2018.

\bibitem{humble}
H. Zhou, ``A humble theory and application for logic encryption,''{\it IACR Cryptology ePrint Archive 2017} pp. 696, [Online]. Available: https://eprint.iacr.org/2017/696.pdf

\bibitem{SFLL}
M. Yasin, {\it et. al.}, ``Provably-Secure logic locking: from theory to practice,'' {\it Proc. of the ACM SIGSAC Conf. on Computer and Communication Security}, pp. 1601-1618, Dallas, Texas, U.S.A., 2017.

\bibitem{ECE}
Y. Shen, A. Rezaei and H. Zhou, ``A comparative investigation of approximate attacks on logic encryptions,'' {\it Asia and South Pacific Design Auto. Conf.}, pp. 271-276, Jeju, 2018.

\bibitem{FALL}
D. Sirone, P. Subramanyan, ``Functional analysis attacks on logic locking,'' {\it Design, Auto. \& Test in Europe Conf. \& Exhibition,} pp. 936-939, Florence, Italy, 2019.

\bibitem{Fangfei}
F. Yang, M. Tang and O. Sinanoglu , ``Stripped functionality logic locking with hamming distance-based restore unit (SFLL-hd) – unlocked,'' {\it IEEE Trans. on Information Forensics and Security,} vol. 14, no. 10, pp. 2778-2786, Oct. 2019.

\bibitem{CAS}
B. Shakya, X. Xu, M. M. Tehranipoor and D. Forte, ``CAS-Lock: a security-corruptibility trade-off resilient logic locking scheme,'' {\it IACR Transactions on Cryptographic Hardware and Embedded Systems,} vol. 2020, issue 1, pp. 175-202.

\bibitem{Florida}
Netlist encryption and obfuscation suite, https://bitbucket.org/kavehshm/neos/src/master/.

\bibitem{Bypass}
X. Xu, B. Shakya, M. M. Tehranipoor and D. Forte, ``Novel bypass attack and BDD-based tradeoff analysis against all known logic locking attacks,'' {\it IACR Cryptology ePrint Archive}, 2017.

\bibitem{CAS-unlock}
A. Sengupta and O. Sinanoglu, ``CAS-Unlock: Unlocking CAS-Lock without Access to a Reverse-Engineered Netlist,'' {\it IACR Cryptology ePrint Archive}, Report 2019.

\bibitem{DCAS-unlock}
S. Bicky, X. Xu, M. M. Tehranipoor, and D. Forte, ``Defeating CAS-Unlock,” {\it IACR Cryptology ePrint Archive}, Report.


\end{thebibliography}
\end{document}